\documentstyle[epsf]{elsart}
\begin{document}

\def\note#1{{\bf[#1]}}

\def\dirac{{\bf \rm D}\!\!\!\!/\,}
\def\wilson{{\bf \rm W}}
\def\ham{{\bf \rm H}}
\def\mbham{{\cal H}}
\def\bmat{{\bf \rm B}}
\def\cmat{{\bf \rm C}}

\begin{frontmatter}

\begin{flushright}
{\normalsize FSU-SCRI-98-15}\\
{\normalsize hep-lat/9802016}\\
\end{flushright}

\title{ 
Spectral flow, condensate and topology in lattice QCD
}
\author{
Robert G. Edwards, Urs M. Heller and Rajamani Narayanan}
\address{
SCRI, The Florida State University, 
Tallahassee, FL 32306-4130, USA}

\begin{abstract}
We study the spectral flow of the Wilson-Dirac operator $\ham(m)$ with
and without an additional Sheikholeslami-Wohlert (SW) term on a
variety of SU(3) lattice gauge field ensembles in the range $0\le m
\le 2$. We have used ensembles generated from the Wilson gauge action,
an improved gauge action, and several two-flavor dynamical quark
ensembles.  Two regions in $m$ provide a generic characterization of
the spectrum. In region I defined by $m\le m_1$, the spectrum has a
gap. In region II defined by $m_1\le m \le 2$, the gap
is closed. The level crossings in $\ham(m)$ that occur in region II
correspond to localized eigenmodes and the localization size decreases
monotonically with the crossing point down to a size of about one
lattice spacing. These small modes are unphysical, and we find the
topological susceptibility is relatively stable in the part of region II
where the small modes cross.  We
argue that the lack of a gap in region II is expected to persist in
the infinite volume limit at any gauge coupling.  The presence of a
gap is important for the implementation of domain wall fermions.
\end{abstract}

\end{frontmatter}

{\bf PACS \#:}  11.15.Ha, 11.30.Rd, 11.30.Fs\hfill\break
{Key Words:} Lattice QCD, Wilson fermions, Topology, Condensate.

\section{Introduction}

The continuum Hermitian Euclidean operator $\ham(m) = \gamma_5 \bigl[
\gamma_\mu (\partial_\mu + iA_\mu) - m \Bigr]$ has a gap of $2|m|$
for $m\ne 0$, i.e., there are no eigenvalues in the region
$(-|m|,|m|)$. Levels crossing zero (level crossings for short) 
in the spectral flow of $\ham(m)$ can occur
only at $m=0$ and the net number of level crossings is related to the
topological charge of the background gauge field $A_\mu$.  Here $m <
0$ describes physical quarks\footnote{We use an unconventional choice
of the sign of the mass parameter $m$. The choice is motivated since
we will only be concerned with $m>0$.}  and the spectral gap obtained
as an ensemble average in the path integral for QCD could be used as a
definition for the physical quark mass.  In addition to the level
crossings that occur at $m=0$, the spectral distribution
$\rho(\lambda;m)$ could have a non-zero value for $\rho(0;0)$
resulting in a spontaneous breakdown of chiral symmetry.

On the lattice we regularize $\ham(m)$ using Wilson fermions:
$\ham_L(m)=\gamma_5\wilson(-m)$  where $\wilson(m)$ is the standard
Wilson-Dirac operator. In the chiral representation of the (Euclidean)
Dirac matrices the structure is
\begin{equation}
\ham_L(m)=\pmatrix{ \bmat - m     & \cmat \cr
                    \cmat^\dagger & -\bmat + m \cr }
\label{eq:H_L}
\end{equation}
with $\cmat$ the discretization of the continuum chiral Dirac operator and
$\bmat$ the Wilson term including, if wanted, the SW improvement term,
and $m$ is a dimensionless parameter. Free massless fermions correspond to
$m=0$ and the (lifted) lattice doublers become massless at $m=2, 4, 6$ and
8. $\ham_L(m)$ plays a central role in one formulation of chiral fermions
on the lattice~\cite{over} and the chiral determinant in this formalism is
defined with a fixed value of $m_0$ in the region $0 < m_0 < 2$. Level
crossings in $\ham_L(m)$ can occur for any value of 
m in the region $0 \le m \le 8$
\cite{top,su2_inst}. Fermion number is violated by an amount equal to the
net number of level crossings in $\ham_L(m)$ that have occurred in the
region $0 \le m \le m_0$ and this enables one to associate an index of the
chiral Dirac operator on the lattice~\cite{over}. The interpretation from
the overlap formalism is essential to make this connection~\cite{over} and
enables one to compute fermion number violating processes~\cite{chiral}.

Due to additive renormalization of the mass term in the Wilson-Dirac
operator the level crossings do not start at $m=0$ away from the
continuum limit. Instead the gap closes at a value of $m=m_1$ in the
region $0 \le m_1 \le 2$. The region $m < m_1$ describes physical
quarks with a non-zero mass and usual measurements, such as hadron
spectroscopy, determination of weak matrix elements etc., are done in
this region. The region $m > m_1$ is unphysical in this sense;
however, it is of interest in defining chiral fermions on the
lattice~\cite{over,overh} for both chiral gauge theories and vector
gauge theories with massless quarks in the domain wall formalism and
the overlap formalism. The original domain wall formalism contained
an infinite number of fermions~\cite{kaplan}.  A practical
implementation involves a truncation to a finite number of
fermions~\cite{domain}, and the region $m > m_1$ is of interest in
this context.

Unlike in the continuum, level crossings on the lattice do not occur
at one value of $m$ but in a region of $m$. This spreading, observed
numerically in~\cite{su2_inst,su3_top,su2_top}, is due to the presence
of the Wilson term in $\ham_L(m)$. Since the Wilson term is of ${\cal
O}(a)$ one might expect that in the continuum limit (a) the region of
level crossings shrinks to zero and (b) the starting value $m_1$ goes
to zero. Thus one would expect that at finite lattice spacing $a$ the
spectral distribution of $\ham_L(m)$ on the lattice be characterized
by three regions in the range $0 \le m \le 2$. In region I, $0 \le m <
m_1$, corresponding to positive physical quark mass, the spectrum has
a gap. In region II, $m_1 \le m \le m_2$, the spectrum is gapless. In
region III, $m_2 < m \le 2$, the spectrum has a gap again. One would
also expect that $m_1$ and $m_2$ both go to zero in the continuum
limit. This is indeed the case in two dimensional U(1) gauge theory as
is evident from a study of the spectral flow~\cite{su2_top}. This
scenario is also supported by the study of the Schwinger model using
the overlap formalism~\cite{Schwinger} and the domain wall
formalism~\cite{vranas} where the spectrum of $\ham_L(m)$ was found to
have a gap at the values of $m$ that were used. One should note,
however, that non-trivial gauge fields, the ``instantons'', in two
dimensional U(1) theory have only one size, namely the size of the
box.

Contrary to these expectations an investigation of the spectral flow on  an
ensemble of pure SU(3) gauge field configurations at $\beta=5.7$ on an
$8^3\times 16$ lattice showed the presence of region I with $m_1=1.02$, but
the region II, where the gap is closed, extended all the way up to
$m=2$~\cite{su3_top}. In this paper we extend the study of the spectral
flow to a variety of SU(3) lattice gauge field ensembles at weaker
couplings. We include an ensemble generated with an improved pure gauge
action and several ensembles generated with the feedback from two flavors
of dynamical quarks. We also study the effect of including the SW
improvement term into $\ham_L(m)$. In all cases we find that, in a
sufficiently large volume, once the gap closes at an $m=m_1$, it does not
open up again before $m=2$ where the doubler modes start playing an
important role. Region III seems to be absent. All level crossings
correspond to localized eigenmodes of $\ham_L(m)$ with a size that
decreases monotonically with increase in the value $m$ at which the
crossing occurs.

However, within the gapless region we can distinguish two different regimes.
One, close to the beginning of the gapless region, {\it i.e.}, close to
$m_1$, is characterized by a large density of small eigenvalues, and the
zero modes have a size of several lattice spacings. The other regime has a
lower density of small eigenvalues and the zero modes have a size of order
one to two lattice spacings. The level crossings in this regime do not seem
to affect physical observables. In particular, the topological
susceptibility, obtained from the net number of level crossings between 0
and $m$, remains constant once $m$ is in this second regime.

In the next section we describe in some detail the spectral flows we
obtained and give a characterization of the density of low eigenvalues.
In
section~\ref{topology} we present the relation between the size of zero
modes and the value of $m$ at which the crossings occur and give the
determination of the topological susceptibility. In
section~\ref{domain_wall} we describe the consequences of our findings for
the use of domain wall fermions in numerical simulations. A summary and
conclusions follow in section~\ref{conclusions}.

The connection between level crossings of $\ham_L(m)$ and the topology of
the (background) gauge field has been confirmed on the lattice with
spectral flows in smooth instanton background fields: the level crossings
were consistent with the gauge field topology and the eigenmodes at the
crossing points had the appropriate shape of the expected zero
modes~\cite{over,su2_inst}. Level crossings have also been measured on an
ensemble of SU(2) pure gauge field configurations at $\beta=2.4$ on a
$12^4$ lattice. The resulting distribution of the index was found to be in
agreement with the distribution of topological charge \cite{su2_top}.

A level crossing, {\it i.e.}, a zero eigenvalues of $\ham_L(m)$ at some
value of $m$, corresponds to a zero eigenvalue of $\wilson(-m)$. Since $m$
is an additive constant in $\wilson(-m)$ this corresponds to the existence
of a real eigenvalue for the Wilson-Dirac operator ($\wilson(0)$) on the
lattice, and vice versa. Several papers have recently appeared where
special emphasis is placed on the real eigenvalues of the Wilson-Dirac
operator~\cite{flurry} and where a connection to topology has been
speculated.

Low lying eigenvalues of $\ham_L(m)$ have been investigated recently,
motivated by the need for an efficient algorithm to deal with Wilson-Dirac
fermions on the lattice~\cite{Qspec1,Qspec2}. Some evidence was found in
these papers that the low lying eigenvalues are localized.

\section{Spectral flows and the density of low lying eigenvalues}
\label{spec_flow}

In a previous publication we studied the spectral flow of $\ham_L(m)$ on an
ensemble of pure SU(3) gauge field configurations at $\beta=5.7$ on an
$8^3\times 16$ lattice~\cite{su3_top}. We found that once the gap closes,
it remains closed all the way up to $m=2$. Here we extend this study to a
variety of SU(3) lattice gauge field ensembles at weaker couplings,
including an ensemble with an improved pure gauge action and several
ensembles with two flavors of dynamical fermions. The relevant parameters
for all ensembles are given in Table~\ref{tab:ensembles}. The first one
listed is from Ref.~\cite{su3_top}. For two ensembles, (5) and (7), we used
the ${\cal O}(a)$ improved Wilson-Dirac operator with the nonperturbatively
determined SW coefficient of Ref.~\cite{alpha,scri}.

We use the Ritz functional~\cite{ritz} to obtain the 10 lowest eigenvalues
in the spectrum of $\ham^2_L(m)$, and thus the 10 eigenvalues of
$\ham_L(m)$ closest to zero. In addition to giving accurate estimates for
the low lying eigenvalues, the Ritz functional method also gives the
corresponding eigenvectors. Using these eigenvectors one can use first
order perturbation theory to interpolate the eigenvalues between successive
mass points. This enables one to obtain the spectral flow of $\ham_L(m)$ as
a continuous function in $m$ from the computation at suitably spaced
values.

\begin{figure}
\epsfxsize=7in
\centerline{\epsffile{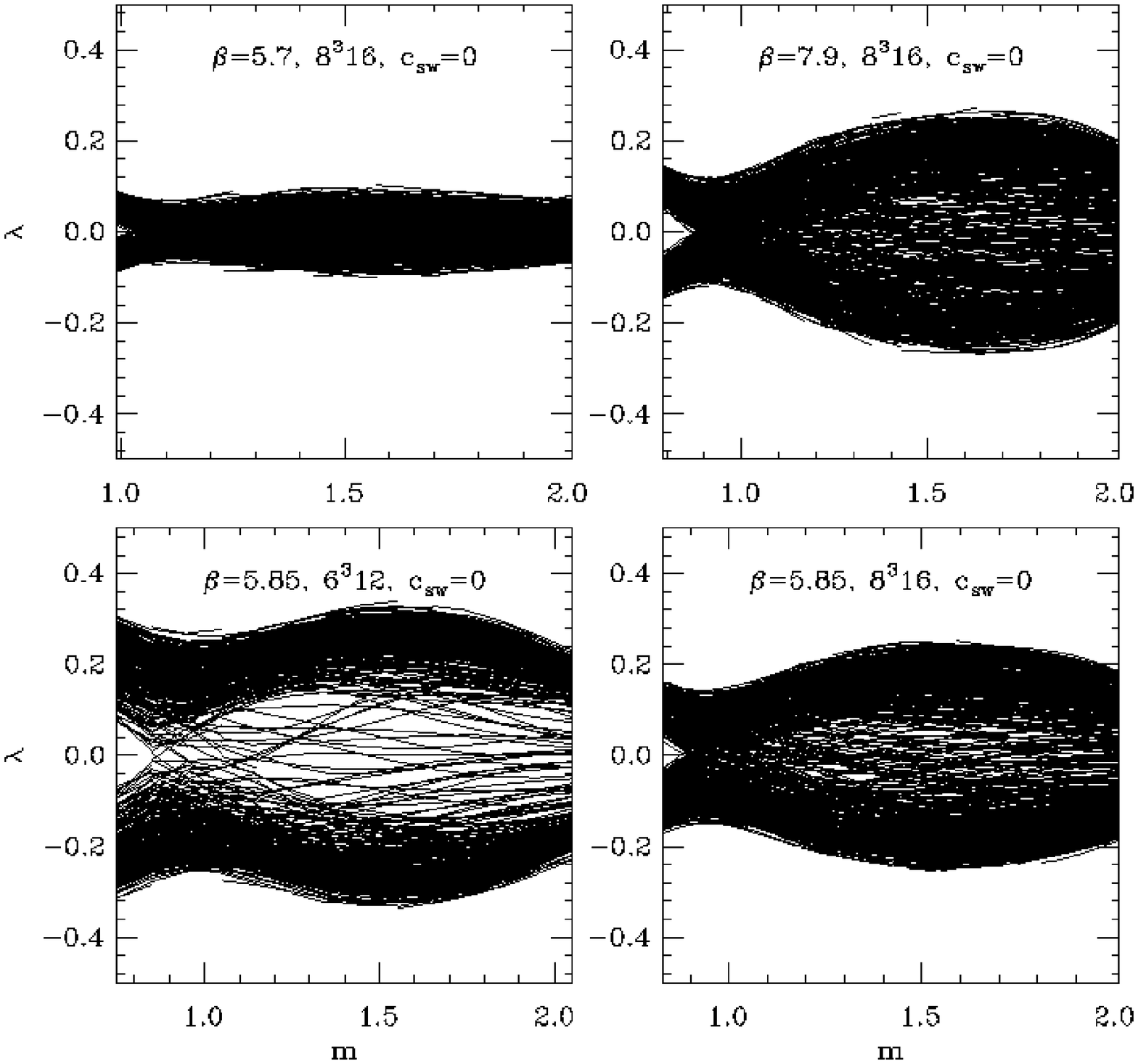}}
\caption{
Spectral flow of the Wilson-Dirac $\ham_L(m)$ 
for the first four quenched ensembles in Table 1.
}
\label{fig:flow_1}
\end{figure}

\begin{figure}
\epsfxsize=7in
\centerline{\epsffile{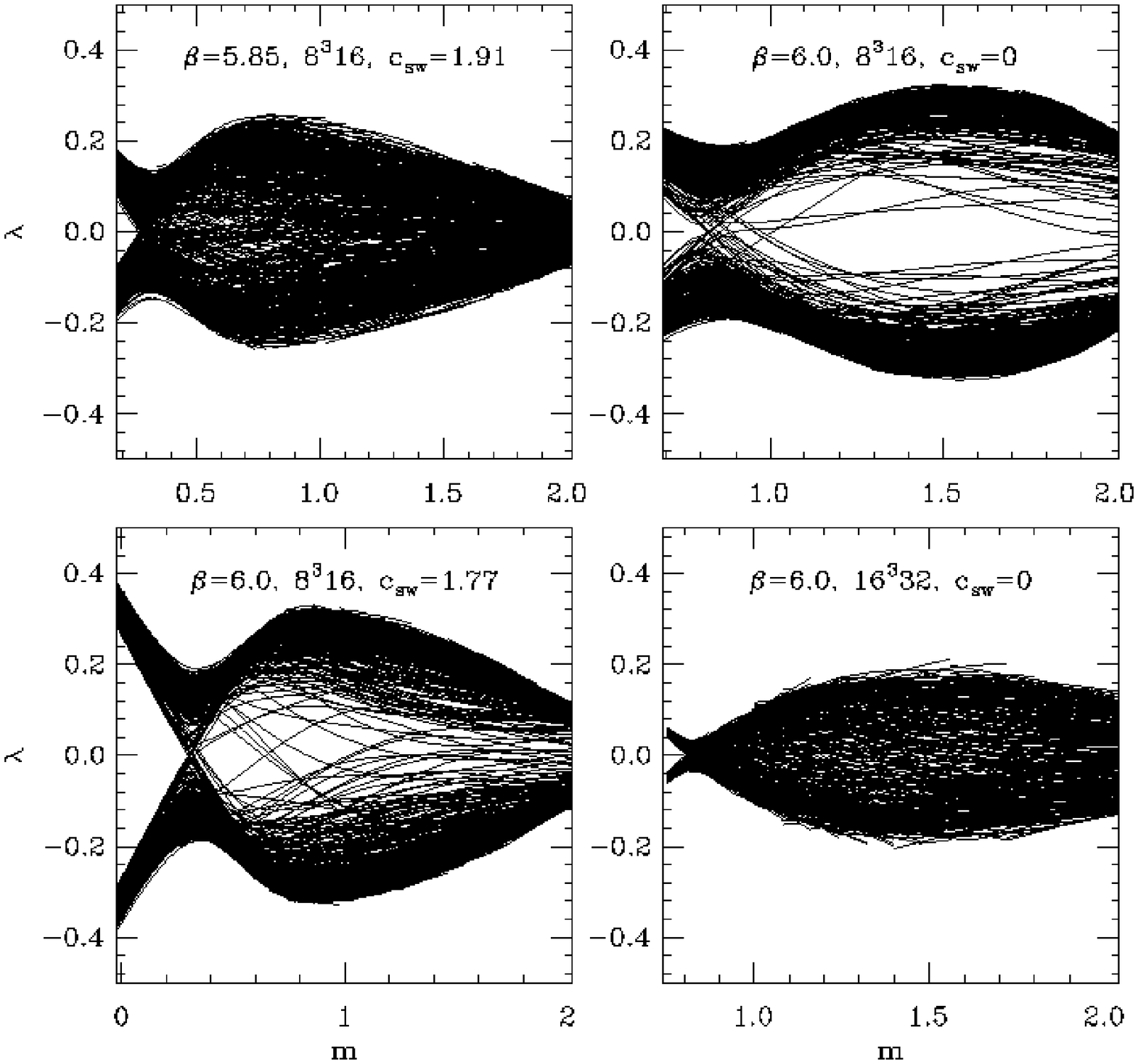}}
\caption{
Spectral flow of the Wilson-Dirac $\ham_L(m)$ 
for the last four quenched ensembles in Table 1.
}
\label{fig:flow_2}
\end{figure}

\begin{figure}
\epsfxsize=7in
\centerline{\epsffile{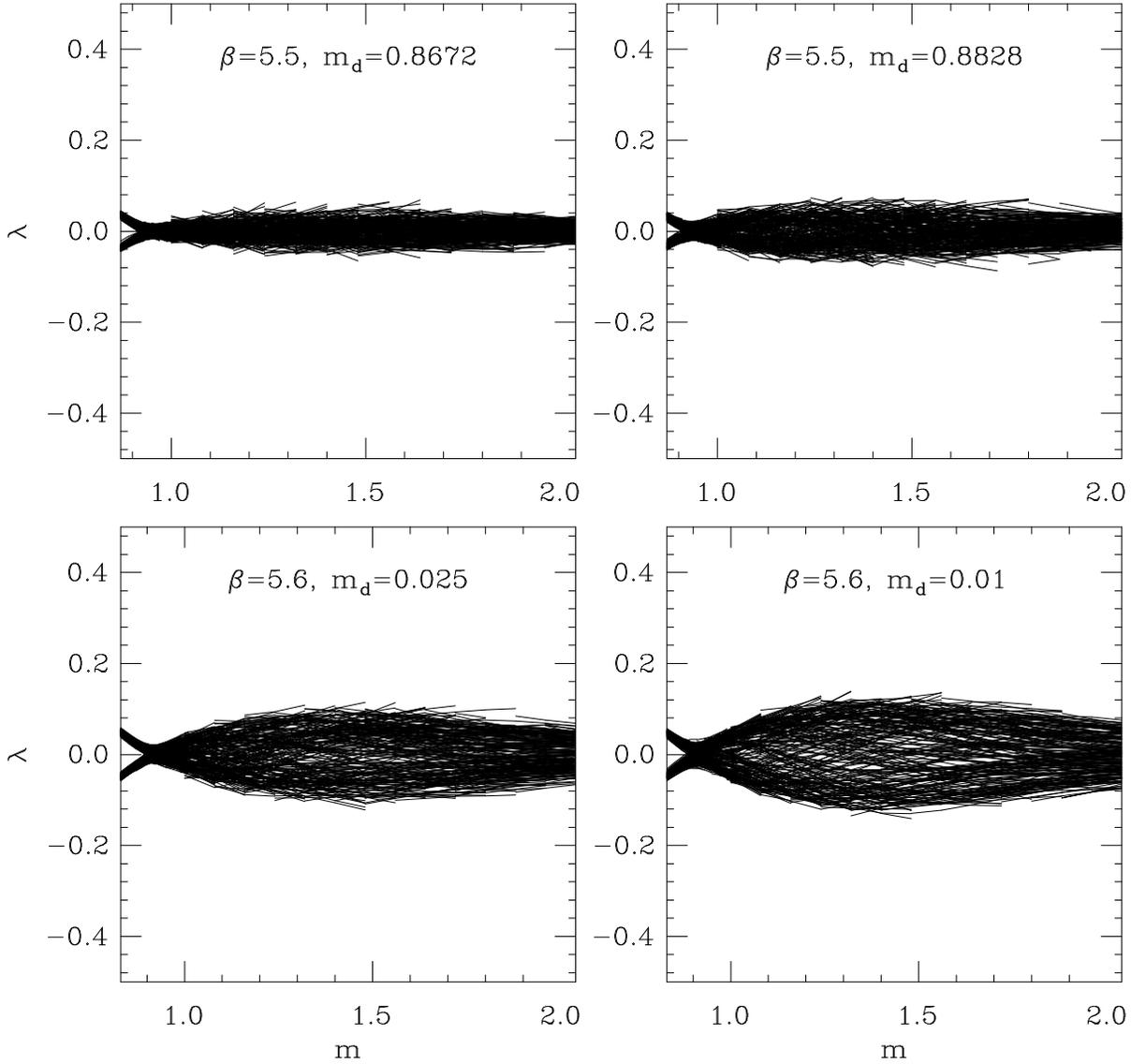}}
\caption{
Spectral flow of the Wilson-Dirac $\ham_L(m)$ 
for four of the dynamical ensembles in Table 1. 
}
\label{fig:flow_3}
\end{figure}

Figures~\ref{fig:flow_1} -- \ref{fig:flow_3} show the cumulative flow for the
various ensembles in Table~\ref{tab:ensembles}. Each figure is a cumulative
plot of the spectral flow of the 10 lowest eigenvalues of $\ham^2_L(m)$ in
one ensemble. Each figure shows a small region of $m$ where the gap is open
-- the end of region I as defined in the introduction. Most of each plot
focuses on what was referred to as region II. This is the region where
level crossings occur in the flow and where there are very low lying
eigenvalues.  The point in $m$ where the first crossing occurs in each
ensemble is by definition the boundary, $m_1$, between regions I and II.
The value of $m_1$ for the various ensembles is quoted in
Table~\ref{tab:closing}. We do not quote an error for this quantity since
it is a number defined for the whole ensemble. The spectral flow for the
ensemble of quenched $\beta=6.0$ configurations on the $8^3\times 16$
lattice using the SW improved $\ham_L(m)$ covers the entire region between
$m=0$ and $m=2$ clearly showing region I where the gap is open.

As expected, $m_1$ decreases as we go to weaker coupling. It also decreases
with the addition of an SW term to $\ham_L(m)$ due to the reduction of the
additive renormalization of the mass term for improved Wilson fermions. One
can compare $m_1$ with $m_c$ for the various ensembles, where $m_c$ is the
``critical'' mass where the pion mass computed in the region of positive
physical quark mass extrapolates to zero. 
We note that where the data exists, 
this definition of $m_c$ agrees well with the PCAC
determination of $m_c$~\cite{alpha}.
For the dynamical configurations,
$m_c$ is defined using spectator Wilson quarks. We find $m_c > m_1$,
indicating that $m_c$ lies inside region II. Increasing statistics, {\it
i.e.} the number of configurations analyzed, can only decrease $m_1$.

We have plotted all the cumulative flows with the same scale on the y-axis
to facilitate a comparison between the different ensembles.  A comparison
of the spectral flows of the unimproved $\ham_L(m)$ for $\beta=5.7, 5.85$,
and 6.0 on the $8^3\times 16$ lattice shows a thinning of the spectrum near
zero for larger values of $m$ as one goes to larger values of $\beta$. In
fact the gap seems to have essentially opened up for larger values of $m$
at $\beta=6.0$ on the $8^3\times 16$ lattice. However, we should keep in mind
that the physical volume is getting smaller as one goes to higher values of
$\beta$ at fixed lattice size. Two pairs of ensembles, (3) and (4) at
$\beta=5.85$ and (6) and (8) at $\beta=6.0$, show us the effect of
increasing the physical volume: the spectrum gets denser around zero for
larger values of $m$ with increasing volume. 
Noting that in two pairs of ensembles, namely (1,8) at 
($\beta=5.7$,$\beta=6.0$) and 
(3,6) at ($\beta=5.85,\beta=6.0$), the physical volume is
essentially the same, we see that the spectrum gets thinner
for a fixed physical volume as we go closer to the continuum limit.

A great deal of attention has recently been given to improved lattice actions. We
therefore also studied the spectral flow of $\ham_L(m)$ on an ensemble, (2)
in Table~\ref{tab:ensembles}, of quenched configurations generated using
the  Symanzik improved gauge action~\cite{gauge_imp} at a lattice gauge
coupling of $\beta=7.9$ on an $8^3\times 16$ lattice. This ensemble has a
lattice spacing in between those with Wilson action at $\beta=5.7$ and
$5.85$ (see Table~\ref{tab:closing}). The spectral flow looks very much
like the one for the standard Wilson gauge action with the density in the
large mass region in between that for the Wilson action ensembles at
$\beta=5.7$ and $\beta=5.85$. Improving the gauge action seems to have
little effect on the spectral flow. We also studied the effect of improving
$\ham_L(m)$ by adding an SW term with the nonperturbatively determined
SW coefficient of Ref.~\cite{alpha,scri}. We have done this for the
quenched ensembles (5) and (7) on the $8^3\times 16$ lattices at
$\beta=5.85$ and $6.0$. In accordance with the smaller additive mass
renormalization we find that the gap closes at a smaller value $m_1$. The
spectrum is denser around zero for larger values of $m$ and the effect of
doubler modes seems to set in earlier -- we attribute the decrease of the
distance between the largest among the 10 lowest eigenvalues, as $m$
approaches 2, to the appearance of doubler modes. Like for the unimproved
$\ham_L(m)$ the density around zero decreases as we go to the weaker
coupling. 

Finally we look at the effect of dynamical fermions on the spectral flow.
We consider three ensembles with dynamical Wilson fermions and two ensembles
with dynamical staggered fermions. All these ensembles have a lattice
spacing that is roughly equivalent to a quenched Wilson action ensemble at
$\beta=6.0$ (see Table~\ref{tab:closing}). One should note that each of the
dynamical ensembles has only 20 configurations.
In contrast, 
all the quenched
ensembles had 50 configurations, with the exception of the
$\beta=6.0$ ensemble on a $16^3\times 32$ lattice which had 30
configurations. The spectral flows for all the
dynamical configurations look more
like the one for the $\beta=5.7$ quenched configurations indicating that
the dynamical configurations behave as though they are on a coarser lattice
than expected. This effect is more pronounced for the ensembles with
dynamical Wilson fermions than for those with staggered quarks. We note
that for the ensembles with dynamical Wilson fermions, as expected, the gap
closes at a larger mass parameter $m_1$ than the $m_d$ of the dynamical
fermions (recall that due to our sign convention a larger mass parameter
corresponds to a smaller physical mass).

A measure of the density of eigenvalues at zero at a fixed value of $m$ can
be obtained by considering the condensate as defined
by
\begin{equation}
\rho(0;m)=\lim_{h\rightarrow 0}  \lim_{V\rightarrow\infty} \rho_h(m)
\qquad
\rho_h(m) = {1\over V} \langle \sum_n {h\over \lambda^2_n(m) + h^2} \rangle ,
\label{eq:chiral}
\end{equation}
where $<\dots>$ indicates an average over a gauge field ensemble.
In the thermodynamic limit indicated, $\rho(0;m)$ is the density of
eigenvalues at zero (up to an irrelevant multiplicative factor).
In this paper we approximate $\rho_h(m)$, at given fixed volume,
by summing over the ten
lowest eigenvalues and keeping $h$ fixed at $0.01$. When the gap is closed
$\rho_h(m)$ of (\ref{eq:chiral}) is dominated by the lowest few
eigenvalues. Our aim is to get a qualitative comparison of the dimensionless
ratio $\rho_h(m)/\sigma^{3/2}$ for the various ensembles in
Table~\ref{tab:ensembles} and to get a better understanding of the region
where the gap is closed. 

\begin{figure}
\epsfxsize=7in
\centerline{\epsffile{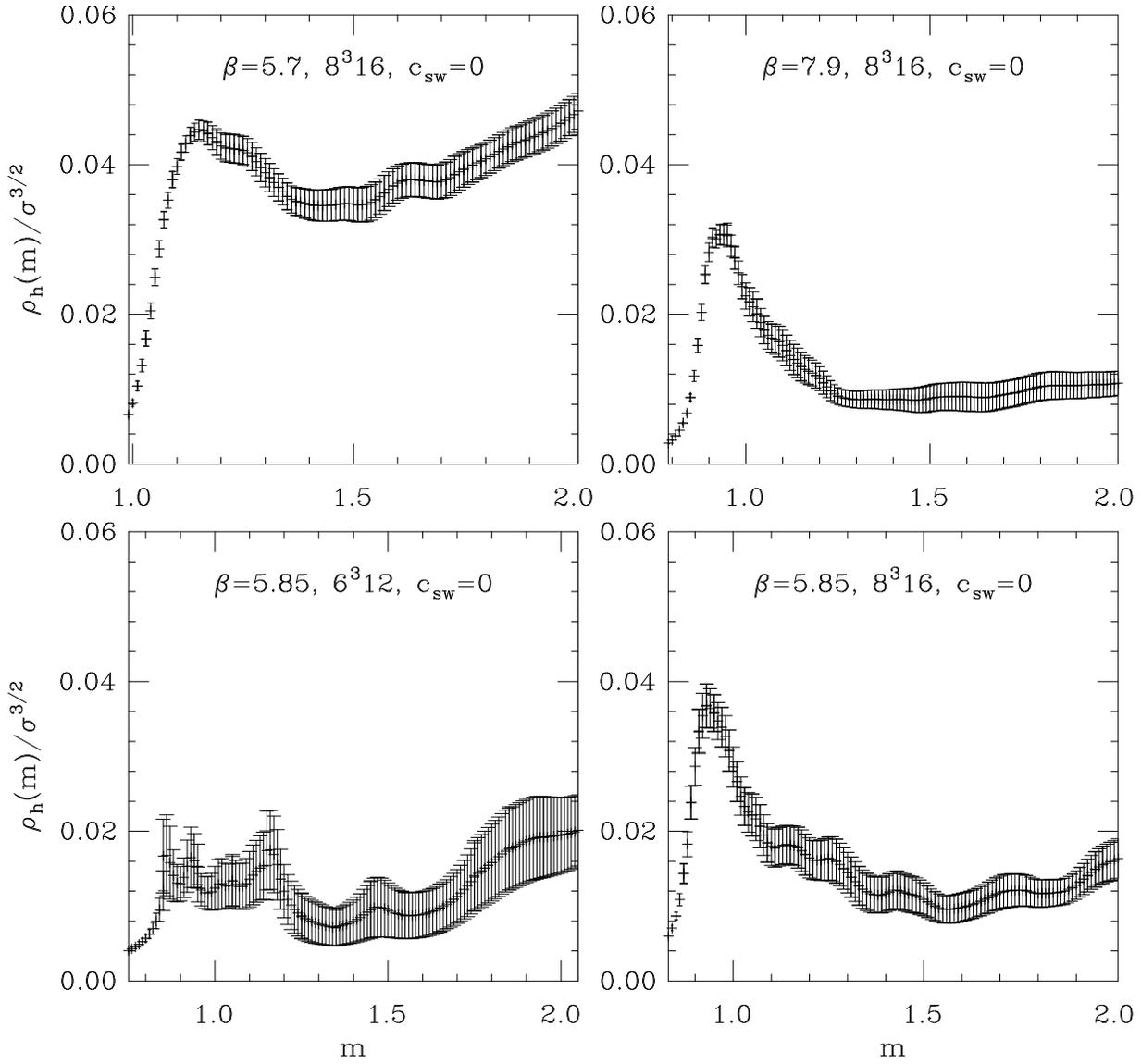}}
\caption{
A measure of the condensate 
for first of four quenched ensembles in Table 1.
}
\label{fig:chiral_1}
\end{figure}

\begin{figure}
\epsfxsize=7in
\centerline{\epsffile{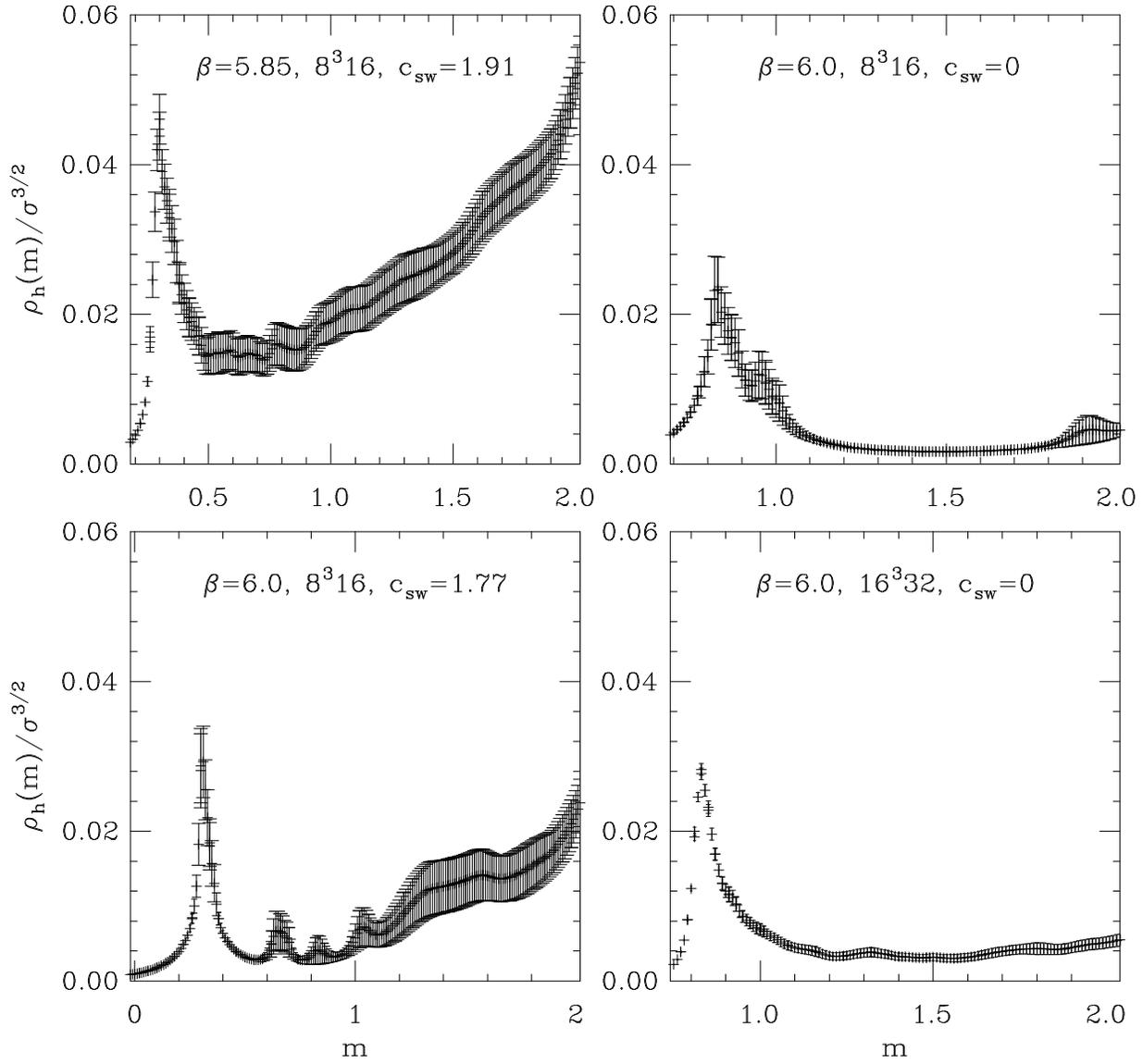}}
\caption{
A measure of the condensate 
for the last four quenched ensembles in Table 1.
}
\label{fig:chiral_2}
\end{figure}

\begin{figure}
\epsfxsize=7in
\centerline{\epsffile{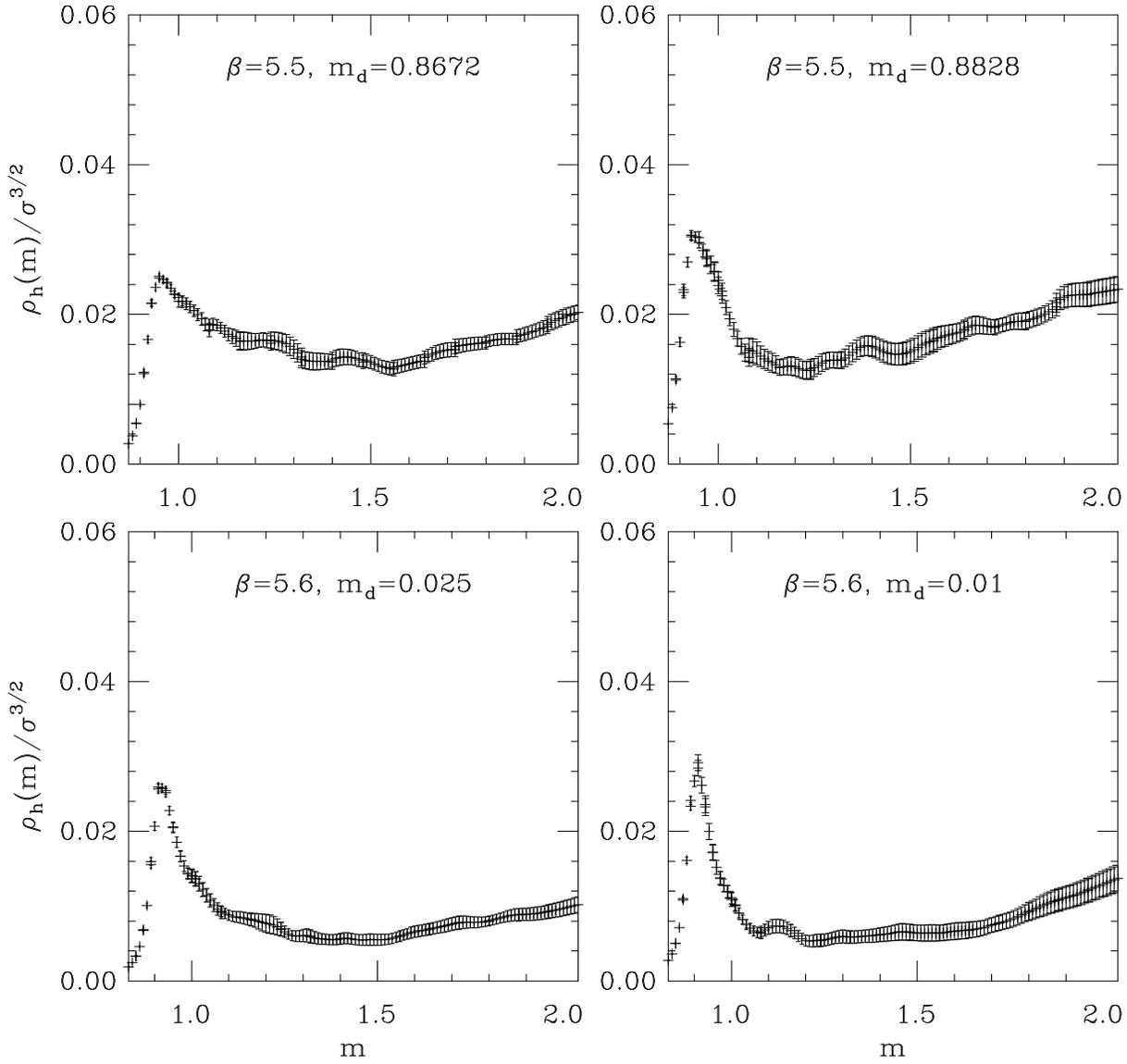}}
\caption{
A measure of the condensate 
for four of the dynamical ensembles in Table 1.
}
\label{fig:chiral_3}
\end{figure}

The various plots are shown in Figures~\ref{fig:chiral_1} --
\ref{fig:chiral_3}. Comparing the plots for $\beta=5.7, 5.85$ and $6.0$ on
the $8^3\times 16$ lattice with no improvement term in $\ham_L(m)$, we see
that a peak develops in the condensate close to the lower end of
region II where the gap is closed. The peak becomes sharper as we go to
weaker coupling and the condensate at larger values of $m$ becomes smaller.
The addition of the SW term to $\ham_L(m)$ has the effect of raising the
peak, but we do not see clear evidence for a sharpening of the peak.
The suppression of the condensate at larger values of $m$, seen
without the SW term, is now weakened. We see a rise in the condensate as
$m$ is increased towards 2. This rise becomes smaller for weaker coupling
albeit also in a smaller physical volume. We attribute this rise to an
increased density of low eigenvalues due to doubler modes starting to
become small.

Improving the gauge action ($\beta=7.9$ on an $8^3\times 16$ lattice) does
not change the qualitative behavior of the condensate when compared
to a standard Wilson action ensemble. However, the condensate for larger values
of $m$ seem to be smaller than the condensate obtained using the standard
Wilson action at $\beta=5.85$ even though the Symanzik improved ensemble is
thought to be on a coarser lattice as seen comparing the values of the
string tension in Table~\ref{tab:closing}. 

Turning now to the dynamical ensembles, we see a peak in the 
condensate and a slow rise for larger values of $m$. The peak seems to
become somewhat sharper as one reduces the dynamical quark mass indicating
a stronger signal for a condensate. The rise at larger mass
parameter $m$ indicates that these configurations are farther away from the
continuum limit than the quenched configurations at $\beta=6.0$, with the
dynamical Wilson ensembles being farthest away.

\section{Topology from level crossings}
\label{topology}

\begin{figure}
\epsfxsize=7in
\centerline{\epsffile{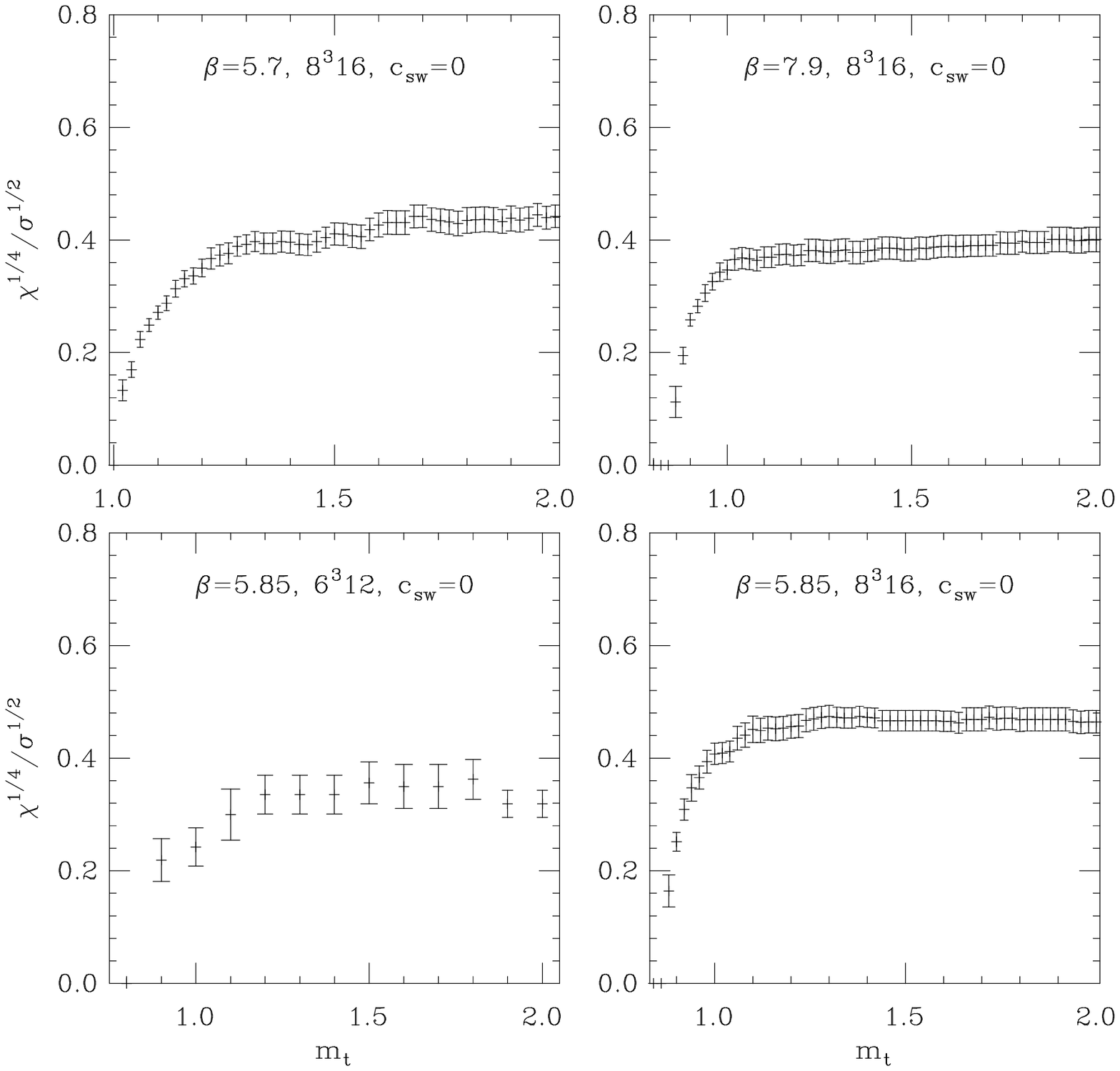}}
\caption{
Topological susceptibility as a function of $m$ 
for the first four quenched ensembles in Table 1.
}
\label{fig:chi_1}
\end{figure}

\begin{figure}
\epsfxsize=7in
\centerline{\epsffile{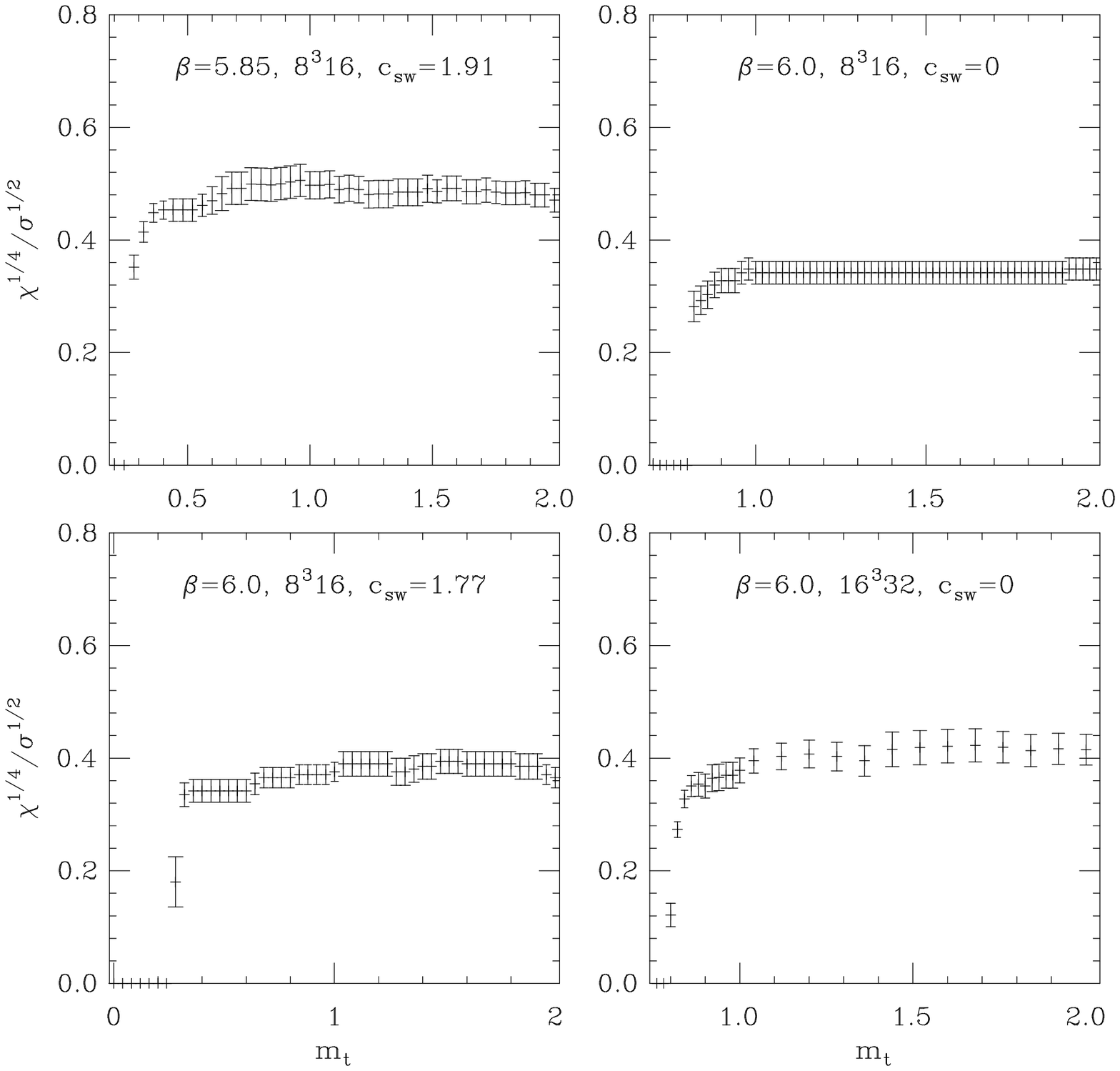}}
\caption{
Topological susceptibility as a function of $m$ 
for the last four quenched ensembles in Table 1.
}
\label{fig:chi_2}
\end{figure}

\begin{figure}
\epsfxsize=7in
\centerline{\epsffile{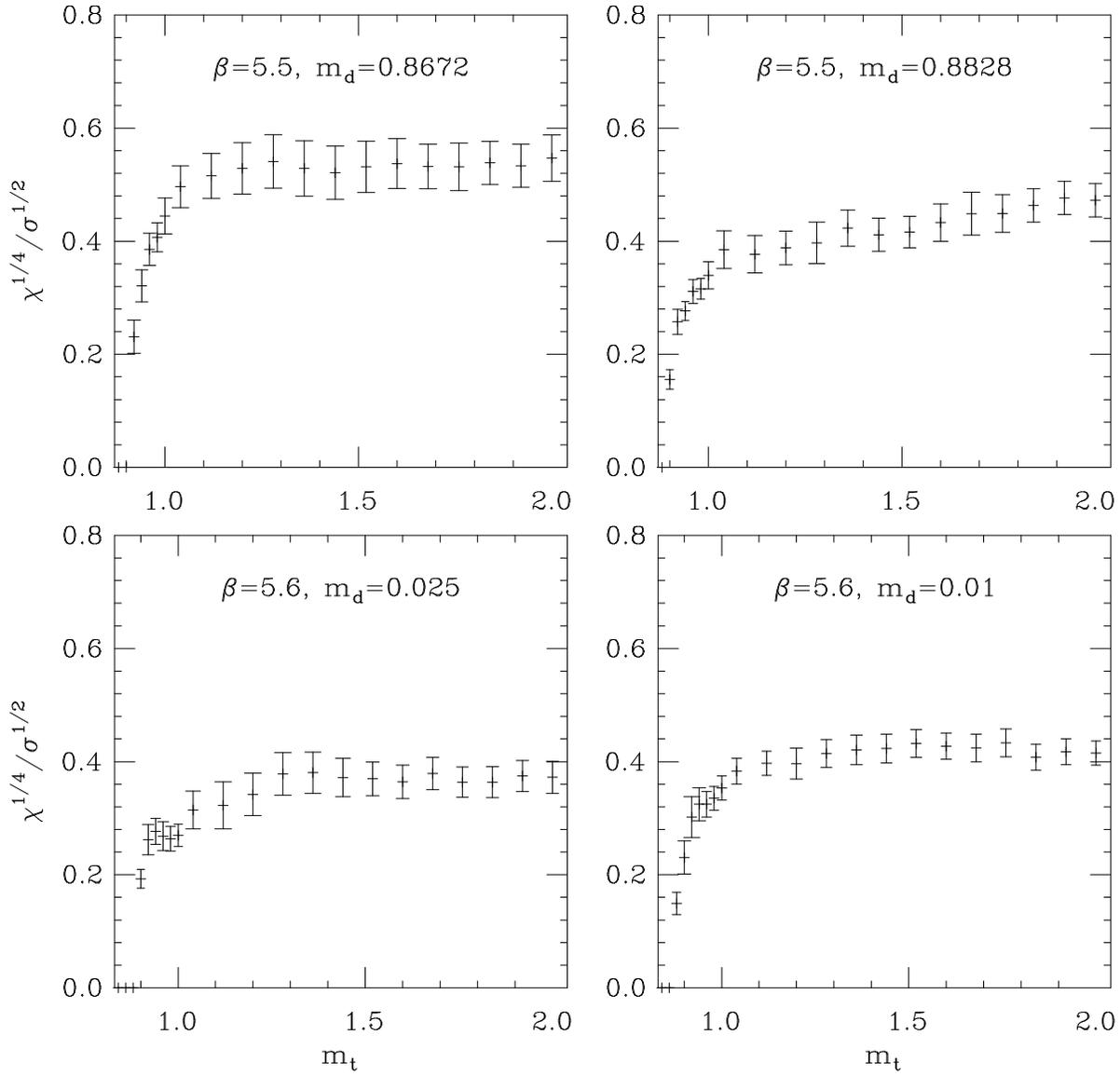}}
\caption{
Topological susceptibility as a function of $m$ 
for four of the dynamical ensembles in Table 1.
}
\label{fig:chi_3}
\end{figure}

The net number of level crossings in $\ham_L(m)$ between $m=0$ and a fixed
value $m_t < 2$ results in a distribution of the index of the chiral Dirac
operator at the chosen value $m_t$. Assuming that the index is the same as
the topological charge of the gauge field background we can get an estimate
for the topological susceptibility as a function of $m_t$. Naturally it can
be non-zero only in region II since level crossings occur only in this
region. By investigating the topological susceptibility as a function of
$m_t$ we can get some understanding of the physical relevance of the
crossings that occur at the larger values of $m$ where the 
condensate is small. The various plots are shown in 
Figures~\ref{fig:chi_1} -- \ref{fig:chi_3}. All of them have the general
characteristic that they show a sharp rise in the region where the 
condensate is peaked and then they essentially flatten out in the region
where the condensate is small. This is the case also in the presence
of the SW term, for the improved gauge action and for the ensembles that
contain dynamical fermions. This presents convincing evidence that the
level crossings that occur at larger $m$ are not physically relevant.

Further supporting evidence can be found by looking at the eigenvectors
associated with level crossings. We find that all these eigenvectors are
well localized on the lattice. To characterize the localization we use a
definition of the  size of the eigenvector inspired by the 't Hooft zero
mode and given in Ref.~\cite{su3_top}. Other definitions 
motivated by an Anderson type localization can be found
in~\cite{Qspec2}. However, since there is a natural connection between level
crossings in $\ham_L(m)$ and topology, and since for smooth instanton
backgrounds the shape of the eigenmode at a crossing point agrees very well
with the 't Hooft zero mode~\cite{su2_inst}, we prefer the definition in
Ref.~\cite{su3_top}. We should emphasize that we look only at the sizes of
eigenmodes that cross, and only close to the crossing point. Only then can
we expect to get a good estimate of the localization size inspired by the
't Hooft zero mode.

\begin{figure}
\epsfxsize=7in
\centerline{\epsffile{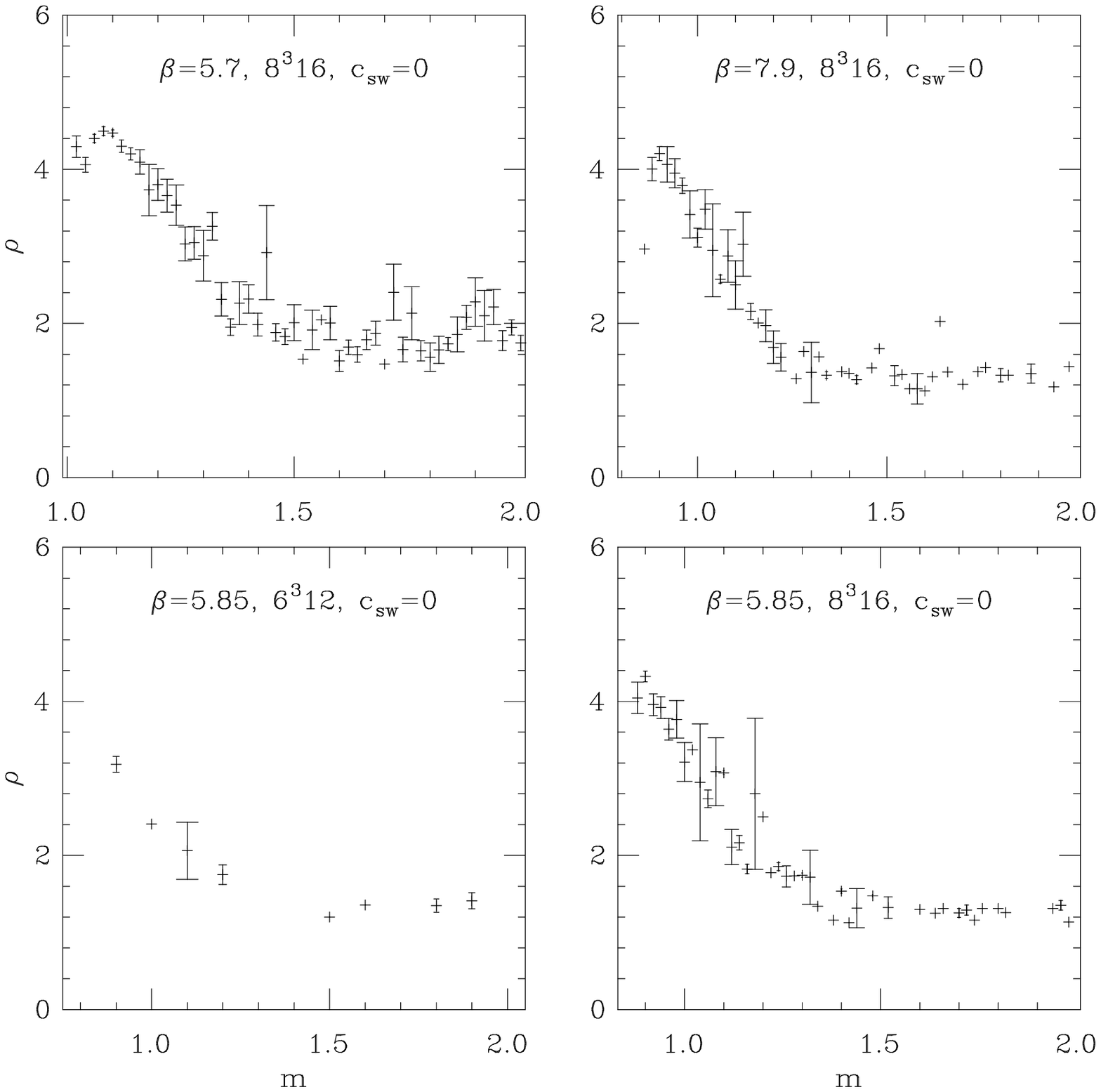}}
\caption{
Localization size of crossing mode in lattice units as a
function of the crossing point for the first four quenched
ensembles in Table 1.
}
\label{fig:rho_1}
\end{figure}

\begin{figure}
\epsfxsize=7in
\centerline{\epsffile{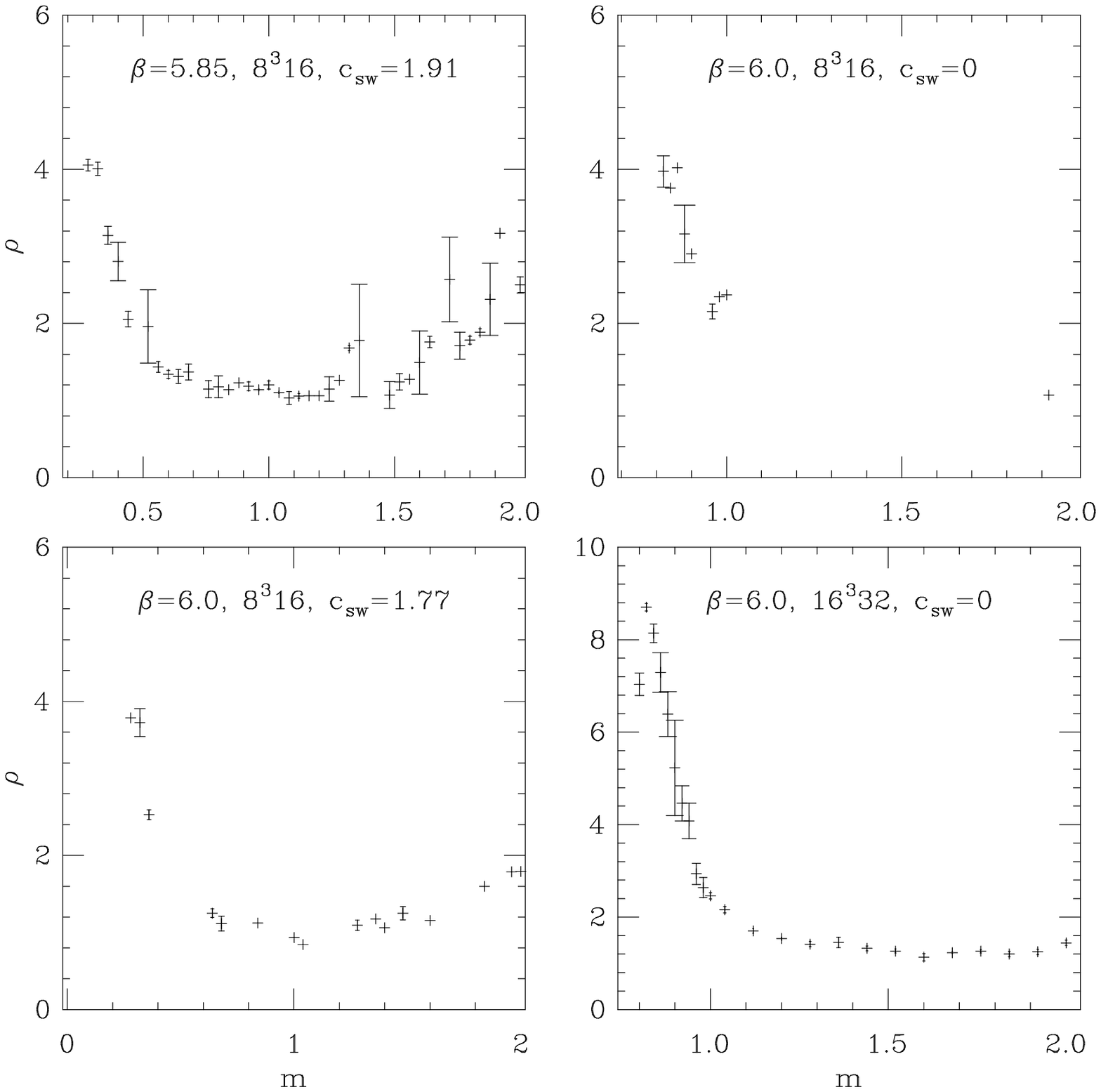}}
\caption{
Localization size of crossing mode in lattice units as a
function of the crossing point for the last four quenched
ensembles in Table 1.
}
\label{fig:rho_2}
\end{figure}

\begin{figure}
\epsfxsize=7in
\centerline{\epsffile{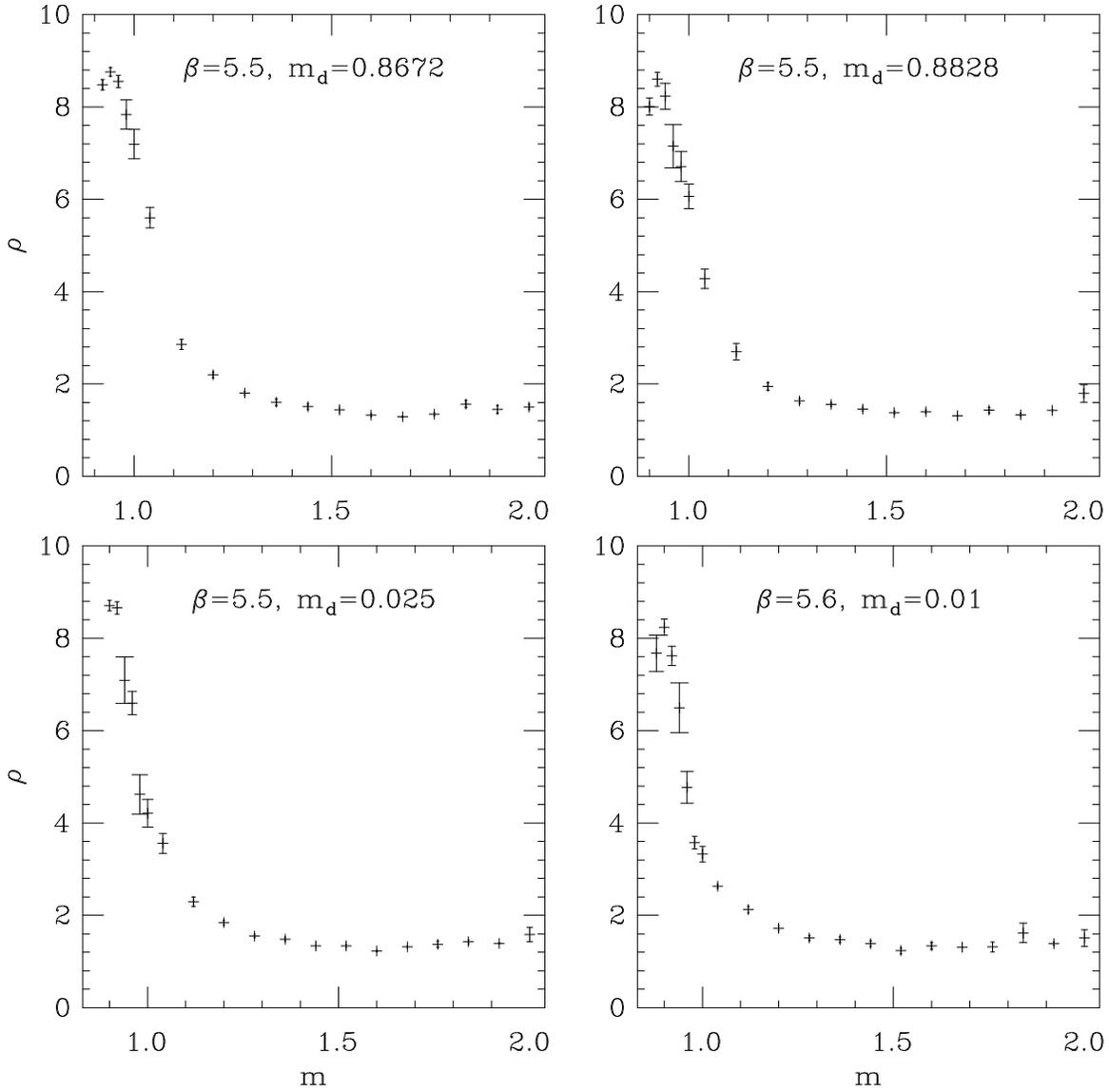}}
\caption{
Localization size of crossing mode in lattice units as a
function of the crossing point for four of the dynamical
ensembles in Table 1.
}
\label{fig:rho_3}
\end{figure}

For each of the ensembles in Table~\ref{tab:ensembles}, we have measured
the sizes of all the modes that cross and we plot the localization size, in
units of the lattice spacing, as a function of the crossing point $m$ in
Figures~\ref{fig:rho_1} -- \ref{fig:rho_3}. All plots show a monotonic
decrease in the localization size as a function of the crossing point.
There is a small deviation from the monotonic decrease at large $m$ for the
flows obtained with an SW improvement term, probably due to the onset of
crossings of doubler modes. The decrease of size is faster as one goes
closer to the continuum limit. We can not see a significant change in the rate of
decrease when we compare the plots with and without the SW term in
$\ham_L(m)$. In all cases we see that the localization sizes associated
with the level crossings in the region where the condensate is small
are of the order of one to two lattice spacing. This shows that they are
not physical, consistent with the finding that these crossings do not
affect the topological susceptibility.

\begin{figure}
\epsfxsize=7in
\centerline{\epsffile{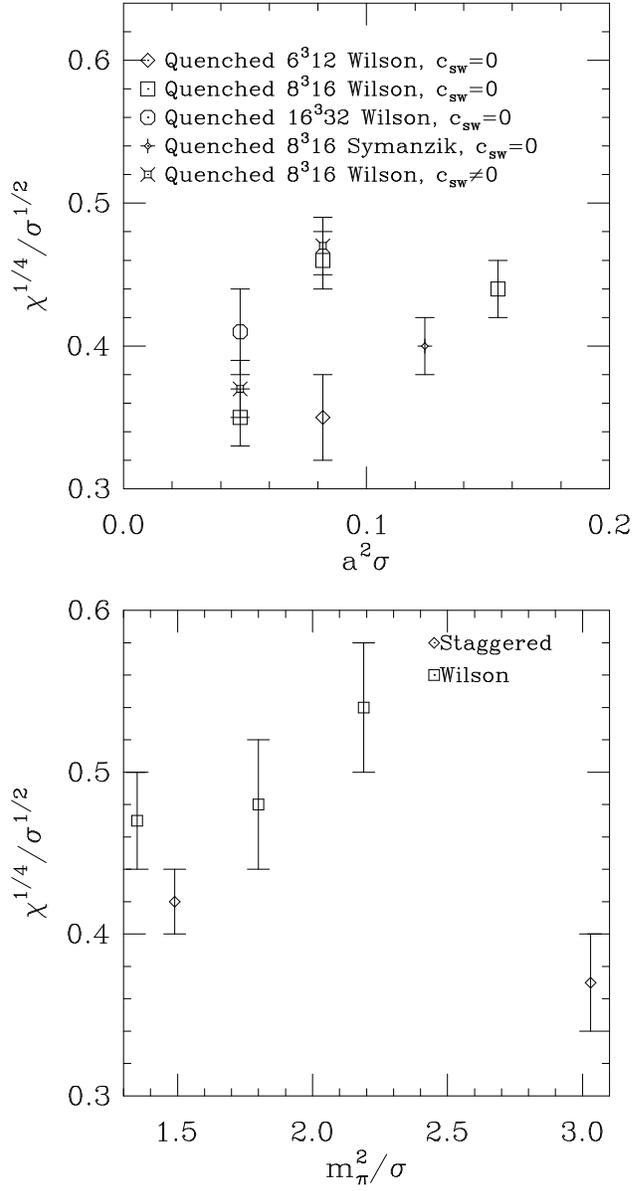}}
\caption{
Plots of the topological susceptibility for the eight quenched
ensembles and the five dynamical ensembles as a function of
the string tension.
}
\label{fig:chi}
\end{figure}

The estimate for the topological susceptibility from the flat region for
the various ensembles, measured in units of the corresponding string
tension, is listed in Table~\ref{tab:closing} and shown in
Figure~\ref{fig:chi}. The result for the quenched ensembles are plotted
together as a function of $a^2\sigma$ to show the trend toward the
continuum limit. There is no hard evidence for scaling yet though we do see
that the results with and without the SW term agree. We note that the
topological susceptibility at $\beta=5.85$ on the $6^3\times 12$ lattice
and at $\beta=6.0$ on the $8^3\times 16$ lattice suffer from finite volume
effects. For the dynamical ensembles, we plot the topological
susceptibility as a function of $m^2_\pi/\sigma$, a
measure of the physical quark mass. 
We note that the topological
susceptibility measured using cooling on the two-flavor staggered
ensembles~\cite{DK} agree with the values here.

\section{Consequences for domain wall fermions}
\label{domain_wall}

Our study of the spectral flow of $\ham_L(m)$ has consequences to
the study of vector gauge theories with massless quarks 
using domain wall fermions~\cite{domain}.
In the domain wall formalism with no explicit quark mass term,
the fermion mass
induced by the finite extent of the extra direction is exponentially
suppressed for free
fermions~\cite{domain} and in one-loop perturbation
theory~\cite{yamada}. Analytical studies pertaining to the
domain wall formalism in the limit of 
an infinite extent in the extra direction
can be found in Ref.~\cite{domain1}.

The many body Hamiltonian for the
propagation in the extra dimension in the domain wall formalism 
is $a^\dagger \ham_D(m) a$, 
where 
\begin{equation}
e^{-\ham_D(m)}= T(m) = \pmatrix
{{1\over 1-m+\bmat} & {1\over 1-m+\bmat}\cmat \cr
\cmat^\dagger {1\over 1-m+\bmat} & \cmat^\dagger {1\over 1-m+\bmat}\cmat
+1-m+ \bmat \cr}
\end{equation}
is the transfer matrix for propagation in the fifth direction with
the domain wall mass set to $m$~\cite{over,overt}.
In order to have an exponential suppression in the fifth direction
in all physical observables, $\ham_D(m)$ should have a gap.
The extent of the fifth direction, $L_s$, has to be chosen proportional to
the inverse of the gap in  $\ham_D(m)$ to suppress dependence on
the extent of the fifth direction~\cite{overt}.
Zero eigenvalues of $\ham_L(m)$ studied in this paper are in one to
one correspondence with zero eigenvalues of $\ham_D(m)$~\cite{over}.
In addition, the slope of the flow at the crossing point is
the same~\cite{over}. 
Therefore the spectrum of $\ham_D(m)$ in the neighborhood of zero
will be the same as the spectrum of $\ham_L(m)$. In particular, if
$\ham_L(m)$ does not have a gap in some region of $m$, then
$\ham_D(m)$ also will not have a gap in the same region of $m$.
Then the ground state of $a^\dagger \ham_D(m) a$
is not separated from the first excited state by a gap and the
dependence of observables on the extent of the extra direction
will not be exponentially suppressed. This could apply, for example,
to the fermion mass induced by the finite extent, $L_s$. We remark
that the zero modes of $\ham_D(m)$ that are relevant for the discussion
here do not correspond to zero modes of the five dimensional 
domain wall Dirac operator. 

For the
domain wall fermions to feel the effect of the gauge field topology, the
domain wall mass must be larger than $m_1$. Our analysis shows that the gap
does not open up again, and therefore there is no choice for the domain
wall mass that leads to an exponential suppression. This fact is 
indepenent of the shape and sizes of the low lying  modes.
We would like to emphasize that the number of modes near zero
increases with the volume as demonstrated by the condensate in
Figures~\ref{fig:chiral_1} and \ref{fig:chiral_2}.
The actual magnitude of the contamination from
the excited states, on the other hand, will depend on the operator studied 
in the domain wall formalism, size and shape of the low lying modes of
$\ham_D(m)$ and the spectral density of $\ham_D(m)$ near zero.

Domain wall fermions have been studied numerically on four dimensional
quenched ensembles~\cite{blum1,blum2}. However, these studies included an
explicit non-zero positive quark mass, $m_q$. Effects of this
explicit quark mass can be understood by looking
at the fermionic determinant in an external
gauge field background at finite extent $L_s$, given by~\cite{overt}
\begin{equation}
\det \Bigl[ { 1+m_q + (1-m_q) \gamma_5 \tanh ({1\over 2} L_s \ham_D)\over 2}
\Bigr]
\end{equation}
where $m_q \le 1$. 
%In the limit of $m_q=1$, there is no dependence on
%the gauge fields at all. The dependence on the gauge fields gets stronger
%as one approaches $m_q=0$. To obtain a result that is essentially independent
%of $L_s$, one will have to go to larger $L_s$ as one approaches $m_q=0$.
%If there is a gap in $\ham_D$, then the increase in $L_s$ that is
%needed will dependent
%only logarithmically in the decrease in $m_q$. If there is no gap in
%$\ham_D$, the increase in $L_s$ will be linear in the decrease in $m_q$.
Deviation of $\tanh ({1\over 2} L_s \ham_D)$ from $\epsilon(\ham_D)$ gives
an additional, induced mass. If $\ham_D$ has a gap, 
this induced mass is exponentially
suppressed, and to keep it at a constant fraction of $m_q$ would require only
an increase of $L_s$ logarithmic in $m_q$, as $m_q$ is decreased. If there is
no gap, the induced mass decreases only as some power of $1/L_s$ and
$L_s$ has to be increased as some (fractional) power of $1/m_q$.

Evidence in support of the need to increase $L_s$ with decreasing
$m_q$ exists in the data for
the pion mass measured as a function of quark mass at $\beta=5.85$ in
Ref.~\cite{blum3}. With a quark mass of $m_q=0.075$ they find that
it is sufficient to have fourteen slices in the extra direction.
But this is not sufficient at $m_q=0.05$ where the result from eighteen
slices in the extra direction significantly deviates from the result
obtained with fourteen slices. However, the results are not detailed
enough to distinguish between the need of only a logarithmic increase
in $m_q$, or a faster one, as we predict due to the absence of a gap.
We find that the gap also remains closed at $\beta=6.0$ on an
$16^3\times 32$ lattice and here again we predict the need of a faster
than logarithmic increase of $L_s$ with decreasing $m_q$. 

We have seen some evidence that the gap gets thinner for a fixed physical
volume as one goes closer to the continuum. This is based on a comparison
of the $\beta=5.7$ ensemble on the $8^3\times 16$ lattice with the
$\beta=6.0$ ensemble on the $16^3\times 32$ lattice and a comparison
of the $\beta=5.85$ ensemble on the $6^3\times 12$ lattice with the
$\beta=6.0$ ensemble on the $8^3\times 16$ lattice.
It is therefore
conceivable that there exists a large enough physical volume where one does
not expect any finite volume effects and a weak enough lattice coupling
where the gap is essentially open, or at least the density of zero
eigenvalues is small enough that one can use domain wall fermions
efficiently to investigate massless QCD on the lattice. Based on
our results, we 
predict that one needs $\beta > 6.0$
if one wants to work with a short extent in the  extra direction
and be able to deal with very small explicit quark masses.

We have previously argued that the crossings occurring at larger $m$, where
the density of low lying eigenvalues, and hence the condensate, is
small, are unphysical. They do not affect the topological susceptibility
and the localization size of the crossing modes are only of the order of
one to two lattice spacing. It is these crossings, however, which hinder the
use of domain wall fermions to study vector gauge theories. 
These crossings are generic and cannot
be eliminated by any improvement of the Wilson-Dirac operator
$\ham_L(m)$. This claim is based on the following heuristic argument:
$\ham_L(m)$ has the structure of Eq.~(\ref{eq:H_L}), with $\cmat$ a
discretization of the continuum chiral Dirac operator with all improvements
and $\bmat$ the Wilson term, again with all improvements.
As argued in
Ref~\cite{over} the  non-commutativity of $\bmat$ and $\cmat$ is an
essential feature of the generic Wilson-Dirac operator on the lattice that
enables it to correctly describe a single physical particle and also
correctly encode the topological content of the background gauge field. The
operators $\bmat$ and $\cmat$ are bounded on the lattice and therefore the
spectra of $\ham_L(m)$ at $m=\pm\infty$ are identical and contain an equal
number of positive and negative eigenvalues. Therefore all crossings that
occur in $\ham_L(m)$ have to occur in pairs. Due to the presence of the Wilson term
$\bmat$, crossings due to the unphysical particles (the doubler modes) do
not occur in the neighborhood of $m=0$. In the continuum limit and for
arbitrary smooth configurations, we expect crossings near $m=0$ due to the
topological content of the background gauge field. Then we expect crossings
in the opposite direction due to doubler modes near $m=2$. This has been
verified for smooth SU(2) instanton  backgrounds~\cite{su2_inst}. There it
was also shown that the crossing point moved farther away from $m=0$ as the
instanton size was decreased and the gauge configuration on the lattice
thus became rougher. Not surprisingly an instanton with a size of the order
of one lattice spacing crossed farthest away from zero. For instantons
below a certain size (about one lattice spacing) no crossings were found.
Since the crossing near $m=0$ is associated with an opposite crossing near
$m=2$ due to the doubler, and since the flow changes continuously as we
reduce the size of the background instanton, the two crossing points have
to move closer together before they coincide, with the flow being tangent,
and then disappear. This picture has to remain true for any $\ham_L(m)$
that conforms to the structure of Eq.~(\ref{eq:H_L}). Therefore crossings
will occur for any value of $m$ in the region $m_1 \le m \le 2$.  Since the
action of the (small) instanton configurations is finite, they will be
generically present in a lattice ensemble. 
But, all crossings beyond a small range
close to the first crossing at $m_1$ are due to small localized modes and
correspond to fluctuations on the scale of the ultraviolet cutoff.

Since the level crossings that occur for large values of $m$ in region II
are due to localized modes of the order of one lattice spacing it is
conceivable that these zero crossings can be ``lifted''. We now
describe one such modification to the domain wall formalism. The modified many
body Hamiltonian describing the propagation in the extra direction has
the following form:
\begin{equation}
\pmatrix{a^\dagger &  b^\dagger \cr} \pmatrix {\ham_L(m) & m_f +\bmat -m_1 \cr
m_f + \bmat - m_1 & - \ham_L(m) } \pmatrix {a \cr b \cr} 
\end{equation}
The diagonal entries describe the propagation of the left and right handed massless
fermions. 
Here, $m_f$ is the bare fermion mass that couples the left and right handed
fermions. 
The operator $\bmat$ is the usual Wilson term also present in $\ham_L(m)$ but is now
used to ``lift'' the zero crossings that occur away from $m_1$ in region II.
In the absence of the off diagonal terms, the spectral flow is a 
superposition of the flow of $+\ham_L(m)$ and $-\ham_L(m)$. 
With $m_f=0$, $\bmat -m_1$ will be small for crossings that occur close to $m_1$,
namely the physical crossings. As such the pair of crossings that occur near $m_1$
will not be affected much and the low lying eigenvalues will remain low.
However,
 the crossings far away from $m_1$ in region II will result in a large
value for $\bmat -m_1$ and will be lifted up. The spectral flow will have low
lying eigenvalues around $m_1$ and will essentially have a gap beyond $m_1$.
The addition of $\bmat -m_1$ results in a mass term and the massless limit will
no longer occur at $m_f=0$ away from the continuum limit. In spite of this,
the above modification might be useful in practice to study domain wall fermions
at reasonable gauge couplings without having to take $L_s$ to
infinity as $m_f$ goes to zero.

\section {Conclusions}
\label{conclusions}

We studied the spectral flow of $\ham_L(m)$ on several lattice gauge
field ensembles with and without the presence of an SW term in
$\ham_L(m)$.  We found the value $m_1$, $0 \le m_1 < 2$, where the gap
of $\ham_L(m)$ closes. In the large volume limit the gap remains
closed for all values $m>m_1$, all the way up to $m=2$. We find $m_1$
decreases when going closer to the continuum limit and also when
adding an SW improvement term to $\ham_L(m)$. In all cases we found
that $m_1 < m_c$ where $m_c$ is the point where the pion mass measured
at values $m < m_1$ extrapolates to zero.  A careful analysis of the
spectral flow using a measure for the condensate showed that
it exhibits a peak in region II where the gap is
closed. The peak occurred at a value very close to $m_1$. The peak became
sharper as one went to weaker coupling; however, we did not see any
significant evidence for a sharpening of the peak upon the addition of the SW
term. The presence of an SW term resulted in a rise in the
condensate for large $m$ and so did the presence of dynamical
fermions. The topological susceptibility measured using the net level
crossing in $\ham_L(m)$ showed a sharp rise at the beginning of region II
and then it remained quite flat for large $m$ where the 
condensate was small.  By studying the size of the eigenmodes that
cross we found that the levels that cross at larger $m$ correspond to
modes that have a localization size of the order of one lattice
spacing.

Improving the gauge action had no dramatic effect on the spectral
flow. The condensate in the large $m$ region showed a slightly
less pronounced rise than the corresponding Wilson gauge results at
$\beta=5.85$ and $5.70$. Some effects of improvement did appear in the
localization size distribution.  In particular, the density of small
modes appears reduced using the improved gauge action while the large
modes near $m_1$ appear unaffected. Both results are within our
expectations from the behavior of the condensate and the action
improvement program.

The spectral flow on a variety of quenched ensembles has shown that
the gap does not open after it has closed at a value of $m=m_1$.
This hinders the use of domain wall fermions in quenched QCD.
A modification of the domain wall fermions to overcome this problem
has been proposed.
One could also study dynamical domain wall
fermions or use them as spectator fermions in dynamical ensembles
generated using Wilson or
staggered fermions. Our study of the flow in this
paper shows that $\beta=5.5$ with dynamical Wilson fermions
and $\beta=5.6$ with dynamical staggered  fermions is not suitable
for a study using domain wall fermions.
It is necessary to go to weaker couplings when dynamical fermions
are present. Our study of the flow of $\ham_L(m)$ has shown that
an addition of an SW term to $\ham_L(m)$ does not help the situation
for domain wall fermions. In fact, it seems to worsen.
On the other hand it is probably not appropriate to include an
SW term in $\ham_L(m)$ in the context of the domain wall formalism since
domain wall fermions with a Wilson-Dirac Hamiltonian are already
expected to be an improvement over Wilson fermions~\cite{yamada,blum2,blum3}.
A study of the spectral flow at weaker couplings than the ones
studied in this paper should help answer the question of
the feasibility of domain wall fermions. 
In ensembles
with dynamical Wilson fermions, only the gap for the dynamical
fermion mass is really relevant for spectroscopy. In all the
cases we studied in this paper, the spectrum has a gap at the
dynamical fermion mass, that is to say that all these correspond
to positive physical quark mass. It would be interesting to study the spectrum
on dynamical ensembles generated with an unphysical quark mass.
There will be a technical obstacle to this study on large volumes
but a small volume study might be feasible.

\ack{
This research was supported by DOE contracts 
DE-FG05-85ER250000 and DE-FG05-96ER40979.
Computations were performed on the QCDSP, CM-2 and the 
workstation cluster at SCRI.}

\begin{table}
\caption{Description of ensembles analyzed in this paper.}
\vskip 0.3cm
\label{tab:ensembles}
\begin{tabular}{|c|c|c|c|c|} \hline
Ensemble & Lattice size & Spectral & 
Flow & Number of \\ 
Type && parameters & parameters & configurations \\ \hline
1) Pure gauge & $8^3\times 16$ & $\beta=5.7$ & 
$c_{sw}=0$ & 50 \\ 
Wilson~\cite{su3_top} &&& $1 < m < 2$ & \\ \hline
2) Symanzik & $8^3\times 16$ & $\beta=7.9$ & 
$c_{sw}=0$ & 50 \\ 
Improved &&& $0.8 < m < 2$ & \\\hline
3) Pure gauge & $6^3\times 12$ & $\beta=5.85$ & 
$c_{sw}=0$ & 50 \\ 
Wilson &&& $0.8 < m < 2$ & \\\hline
4) Pure gauge & $8^3\times 16$ & $\beta=5.85$ & 
$c_{sw}=0$ & 50 \\ 
Wilson &&& $0.84 < m < 2$ & \\\hline
5) Pure gauge  & $8^3\times 16$ & $\beta=5.85$ & 
$c_{sw}=1.91~\cite{scri}$ & 50 \\ 
Wilson &&& $0.2 < m < 2$ & \\\hline
6) Pure gauge  & $8^3\times 16$ & $\beta=6.0$ & 
$c_{sw}=0$ & 50 \\ 
Wilson &&& $0.7 < m < 2$ & \\\hline
7) Pure gauge & $8^3\times 16$ & $\beta=6.0$ & 
$c_{sw}=1.77~\cite{alpha,scri}$ & 50 \\ 
Wilson &&& $0 < m < 2$ & \\\hline
8) Pure gauge  & $16^3\times 32$ & $\beta=6.0$ & 
$c_{sw}=0$ & 30 \\ 
Wilson &&& $0.76 < m < 2$ & \\\hline
9) Two flavor & $16^3\times 32$ & $\beta=5.5$
 &  $c_{sw}=0$ & 20 \\ 
Wilson~\cite{wil_conf} &&$m_d=0.8672$ &$0.88 < m < 2$&\\ \hline 
10) Two flavor  & $16^3\times 32$ & $\beta=5.5$
 &  $c_{sw}=0$ & 20 \\ 
Wilson~\cite{wil_conf} &&$m_d=0.875$ &$0.88 < m < 2$& \\ \hline
11) Two flavor & $16^3\times 32$ & $\beta=5.5$
 &  $c_{sw}=0$ & 20 \\ 
Wilson~\cite{wil_conf} &&$m_d=0.8828$ &$0.88 < m < 2$& \\ \hline
12) Two flavored & $16^3\times 32$ & $\beta=5.6$
 & $c_{sw}=0$ & 20 \\ 
Staggered~\cite{stag_conf} &&$m_d=0.025$ &$0.84 < m < 2$& \\ \hline
13) Two flavored & $16^3\times 32$ & $\beta=5.6$
 & $c_{sw}=0$ & 20 \\ 
Staggered~\cite{stag_conf} &&$m_d=0.01$ &$0.84 < m < 2$& \\ \hline
\end{tabular}
\end{table}

\begin{table}
\caption{A listing of
$m_1$ where the gap closes and the topological susceptibility
for the various ensembles in Table 1}
\vskip 0.3cm
\label{tab:closing}
\begin{tabular}{|c|c|c|c|c|} \hline
Ensemble & $a\sqrt\sigma$ & $m_c$ & $m_1$ & $\chi^{1/4}/\sqrt\sigma$ 
\\ \hline
1 & 0.392~\cite{quen_scale} & 1.047~\cite{quen_crit} & 1.025~\cite{su3_top}
& 0.44(2) \\ \hline
2 & 0.352~\cite{sym_scale} & 0.891~\cite{sym_scale}  & 0.862
& 0.40(2) \\ \hline
3 & 0.286~\cite{quen_scale} & 0.908~\cite{quen_crit} & 0.857
& 0.35(3) \\ \hline
4 & 0.286~\cite{quen_scale} & 0.908~\cite{quen_crit} & 0.873
& 0.46(2) \\ \hline
5 & 0.286~\cite{quen_scale} & 0.283~\cite{scri} & 0.267
& 0.47(2) \\ \hline
6 & 0.220~\cite{quen_scale} & 0.820~\cite{quen_crit} & 0.818
& 0.35(2) \\ \hline
7 & 0.220~\cite{quen_scale} & 0.302~\cite{alpha} & 0.297
& 0.37(2) \\ \hline
8 & 0.220~\cite{quen_scale} & 0.820~\cite{quen_crit} & 0.805
& 0.41(3) \\ \hline
9 & 0.254~\cite{wil_conf} & 0.929~\cite{wil_conf} & 0.914
& 0.54(4) \\ \hline
10 & 0.241~\cite{wil_conf} & 0.922~\cite{wil_conf} & 0.907
& 0.47(4) \\ \hline
11 & 0.226~\cite{wil_conf} & 0.914~\cite{wil_conf} & 0.904
& 0.37(2) \\ \hline
12 & 0.240~\cite{stag_scale} & 0.900~\cite{s_val_kc} & 0.893 
& 0.38(4) \\ \hline
13 & 0.219~\cite{stag_scale} & 0.894~\cite{s_val_kc} & 0.880 
& 0.41(3) \\ \hline
\end{tabular}
\end{table}





\end{document}